\documentclass[11pt,a4paper]{article}
\pdfoutput=1

\usepackage{bm}
\usepackage{amsmath,amssymb}
\usepackage{cite}

\usepackage{graphicx}
\usepackage{color}
\usepackage{upgreek}

\usepackage{hyperref} 
\hypersetup{colorlinks=true, citecolor=blue, filecolor=black, linkcolor=blue, urlcolor=blue, pdfpagemode=UseNone}

\numberwithin{figure}{section}
\numberwithin{equation}{section}

\newcommand{\be}{\begin{equation}}
\newcommand{\ee}{\end{equation}}
\newcommand{\bea}{\begin{eqnarray}}
\newcommand{\eea}{\end{eqnarray}}
\def\beal#1\eeal{\begin{align}#1\end{align}}   
\def\besp#1\eesp{\begin{multline}#1\end{multline}} 

\usepackage[normalem]{ulem}

\newcommand{\cO}[1]{\mathcal{O}_{#1}}

\newcommand{\Op}{\mathcal{O}}
\newcommand{\ph}{\varphi}
\newcommand{\vp}{\varphi}
\newcommand{\tp}{\tilde{\vp}}

\newcommand{\tg}{\tilde{g}}
\newcommand{\tV}{\tilde{V}}

\newcommand{\Lz}{ {\Lambda_0} }

\newcommand\ie{\textit{i.e.}\ }
\newcommand\eg{\textit{e.g.}\ }
\newcommand\cf{\textit{cf.}\ }

\newcommand{\aka}{{a.k.a.}\ }

\newcommand{\half}{\tfrac{1}{2}}

\newcommand{\nn}{\nonumber}

\newcommand{\cu}[1]{\!#1\!}

\usepackage{datetime2}

\begin{document}

\begin{titlepage}

\begin{center}
{\huge \bf Properties of the linearised functional renormalization group}


\end{center}
\vskip1cm


\begin{center}
{\bf Tim R. Morris}
\end{center}

\begin{center}
{\it STAG Research Centre \& Department of Physics and Astronomy,\\  University of Southampton,
Highfield, Southampton, SO17 1BJ, U.K.}\\
\vspace*{0.3cm}
{\tt   T.R.Morris@soton.ac.uk}
\end{center}

\abstract{Interactions growing slower than a certain exponential of the square of a scalar field, are well behaved when evolved under the functional renormalization group linearised  around the Gaussian fixed point. They satisfy properties usually taken for granted, and reproduce standard perturbative quantisation.
However, ever more challenging effects appear the more interactions grow faster than this. We show explicitly that firstly the  flow no longer splits uniquely into operators of definite scaling dimension; then (linearised) flows to the infrared can  end prematurely in a singularity; and finally new interactions can spontaneously appear at any scale.}



\vskip4cm

\end{titlepage}

\tableofcontents


\section{Introduction}
\label{sec:intro}

In this paper we will be mostly concerned with the functional (\aka exact) renormalization group (RG) linearised around the Gaussian fixed point. One might think that everything is known about such a simple situation. However we show that new effects appear once interactions are allowed to grow as fast as an exponential of the square of the field (for large field). These effects challenge our expectations of the RG.  As the speed of growth is increased, the first effect to appear is that the linearised flow no longer splits uniquely into a sum over eigenoperators (operators of definite scaling dimension). The next effect to appear is that the linearised flows towards the IR (infrared) can end prematurely in a singularity (after which the flow ceases to exist). These cases include linear combinations of the hypothesised Halpern-Huang interactions \cite{HHOrig,HHReply,HH2ndpaper,Percacci:2003jz,Periwal:1995hw,Halpern:1997gn,Branchina:2000jp,Bonanno:2000sy,Gies:2000xr,Altschul:2004yq,Altschul:2004gt,Altschul:2005mu,Gies:2009hq,Huang:2010qn,Huang:2011xg,Huang:2011xha,Pietrykowski:2012nc,Huang:2013zaa,Abhignan:2021lub,Morley:2018omd,Morris:1996nx,Morris:1996xq,Bridle:2016nsu}.
Finally, if interactions growing faster than any exponential of the field-squared are allowed, then the effective action at one point on the flow no longer determines its form at lower scales, even at the linearised level.  New interactions can spontaneously appear at any lower scale.

For clarity and simplicity we focus on the linearised effective potential $V(\vp,\Lambda)$ for a single component scalar field $\vp$ (with standard kinetic term), where $\Lambda$ is the effective cutoff, and choose four (Euclidean) dimensions. The effective action is then simply
\be 
\label{action}
\int\!\! d^4x\, \left\{ \half (\partial_\mu\vp)^2 + V(\vp,\Lambda)\right\}\,.
\ee
However it will be clear that the qualitative conclusions are the same 
in any dimension greater than two,\footnote{In two dimensions the engineering dimension of a scalar field vanishes. The linearised RG leads instead to Sine-Gordan models. This was derived  in \cite{Morris:1994jc}, at $O(\partial^0)$ (which at the linearised level is exact).}
for more general field theories, \eg multi-component scalar fields, and for general local interactions (thus also containing space-time derivatives). 
Although there are different versions of the flow equation \cite{Wilson:1973,Wegner:1972ih,Polchinski:1983gv,Nicoll1977,Wetterich:1992,Morris:1993} and different choices of cutoff profile, at the linearised level they all collapse to the same thing. For the effective potential $V(\vp,\Lambda)$ we have (see \eg \cite{Bridle:2016nsu} or sec. 2 of ref. \cite{Morris:2018mhd}):
\be 
\label{flow}
\Lambda\partial_\Lambda V(\vp,\Lambda) = - \frac{\Lambda^2}{2a^2}\, V''(\vp,\Lambda)\,,
\ee
where prime stands for $\partial_\vp$. At the linearised level there is no anomalous dimension. The only quantum correction is the tadpole integral $\langle \ph(x) \ph(x) \rangle$ of the massless scalar field. The effective ultraviolet (UV) regularised version is $\Lambda^2/2a^2$ (independent of $x$),
where the $\Lambda^2$ dependence is guaranteed by dimensions and the dimensionless parameter $a$ captures the entire dependence on regularisation scheme in this situation.

The point of a flow equation is that it tells us how the effective action changes as we lower the cutoff $\Lambda$. Choosing some initial\footnote{Basically this is the bare potential. See refs. \cite{Manrique:2008zw,Morris:2015oca} for the precise relationship.} potential $V(\vp,\Lz)$ at some high scale $\Lambda\cu=\Lz$ we can follow its evolution at the linearised level as we integrate out all the modes, by solving \eqref{flow}. As $\Lambda\cu\to0$, we recover the physical potential $V(\vp,0)$.

We stress that at the linearised level the flow equation \eqref{flow} is exact if we start with only potential interactions. No other approximation has been applied apart from linearisation. In particular the result should not be confused with use of the so-called Local Potential Approximation \cite{Nicoll:1974zz}. The exact quantum correction arises from a term 
\be \int\!\!d^4x d^4y\, \langle \ph(x) \ph(y) \rangle\,\frac{\delta^2}{\delta \vp(x)\delta \vp(y)}\int\!\!d^4z\, V\!\left(\vp(z),\Lambda\right) = \langle \ph(x) \ph(x) \rangle \int\!\!d^4x\, V''\!\left(\vp(x),\Lambda\right)\,, \ee
so no terms are generated other than corrections to the effective potential. 

Of course one should question if linearisation is appropriate. We will come back to this in the Conclusions, sec. \ref{sec:conclusions}.
For the moment we just note that this is standard practice, being the first step in deriving the eigenoperators (see \eg the reviews \cite{Wilson:1973,Morris:1998, Aoki:2000wm,Bagnuls:2000,Berges:2000ew,Polonyi:2001se,Pawlowski:2005xe,Delamotte:2007pf,Kopietz:2010zz,Rosten:2010vm,Dupuis:2020fhh}). Our exposition follows \cite{Bridle:2016nsu,Morris:2018mhd}.  We recast in dimensionless terms using the cutoff, $\vp =  \tp\,\Lambda$, $V=\tV\Lambda^4$, and then separate variables, leading to a solution
\be 
\label{lambda-v}
\tilde{V}(\tilde{\ph},\Lambda) = \tg(\Lambda) \,\tilde{V}(\tilde{\ph})\,,
\ee
where $\tg$ is the scaled coupling
\be 
\label{tg}
\tg(\Lambda) = \frac{g}{\Lambda^{\lambda}}\,,
\ee
$\lambda$ being the RG eigenvalue and $g$ a constant of dimension $\lambda$. According to \eqref{tg}, this linearised coupling grows, stays constant, or shrinks, as we flow to the infrared, \ie
is relevant, marginal, or irrelevant, depending on whether $\lambda$ is positive, zero, or negative, respectively. The function $\tilde{V}(\tilde{\ph})$ satisfies the eigenoperator equation
\be 
\label{eigen}
\lambda\, \tilde{V}(\tp) +\tilde{\ph}\, \tilde{V}' - 4\, \tilde{V} = \frac{\tilde{V}''}{2{a}^2}\,,
\ee
where prime is now differentiation with respect to $\tp$. 

The general non-singular solution of \eqref{eigen} is a linear combination of the Kummer $M$ functions \cite{AbramowitzStegun,OlverAsymptotics,LebedevSpecialFunctions}:
\be 
\label{Kummers}
\omega^\lambda(\tp) := M\!\left(\frac\lambda2-2,\frac12,a^2\tp^2\right)\,,\qquad \tp\, M\!\left(\frac\lambda2-\frac32,\frac32,a^2\tp^2\right)\,.
\ee
These are in fact entire functions of $\tp$, the first (second) being an even (odd) function of $\tp$. For simplicity we will mostly focus on the even eigenoperators and as in \cite{Bridle:2016nsu}  we call them $\omega^\lambda(\tp)$.  

For $\lambda\cu=4-n$, with $n$ an even (odd) non-negative integer, the first (second) solution is proportional to a Hermite polynomial. We normalise these polynomials as\footnote{Kummer functions are normalised such that $M=1$ at $\tp=0$.}
\be 
\label{On}
\cO{n}(\tp) = H_n(a\tp)/(2a)^n = \tp^n -n(n-1)\tp^{n-2}/4a^2 +\cdots\,.
\ee
The (scaling) dimension of the operator $\cO{n}$ is thus $4-\lambda=n$, coinciding with the engineering dimension $[\vp^n]$ of its top term. The lower powers appear due to the tadpole corrections. At the non-linear level, expanding over these $\cO{n}$ operators reproduces perturbation theory \cite{Wilson:1973,Morris:1998, Aoki:2000wm,Bagnuls:2000,Berges:2000ew,Polonyi:2001se,Pawlowski:2005xe,Delamotte:2007pf,Kopietz:2010zz,Rosten:2010vm,Dupuis:2020fhh,Bridle:2016nsu,Morris:2018mhd}. 

The remaining eigenoperator solutions are the hypothesised Halpern-Huang (HH) interactions \cite{HHOrig,HHReply,HH2ndpaper,Percacci:2003jz,Periwal:1995hw,Halpern:1997gn,Branchina:2000jp,Bonanno:2000sy,Gies:2000xr,Altschul:2004yq,Altschul:2004gt,Altschul:2005mu,Gies:2009hq,Huang:2010qn,Huang:2011xg,Huang:2011xha,Pietrykowski:2012nc,Huang:2013zaa,Abhignan:2021lub,Morley:2018omd,Morris:1996nx,Morris:1996xq,Bridle:2016nsu}. They attract interest especially because the $\lambda\cu>0$ operators appear to offer relevant interactions that would allow genuine interacting continuum limits for scalar fields (such as the Higgs field) in four space-time dimensions. They grow like $\text{e}^{a^2\tp^2}$ for large field. 
More precisely we have for the even operators that asymptotically
\be 
\label{asymp}
\omega^\lambda(\tp) \sim \frac{\sqrt{\pi}}{\Gamma(\lambda/2-2)}\,|\tp|^{\lambda-5}\,{\rm e}^{a^2\tp^2}\,, \qquad{\rm as}\quad \tp\to\pm\infty
\ee
($\lambda\ne 4-2n$, $n$ a non-negative integer). 

It will be useful for us to note that  at $\lambda\cu=5+n$ ($n$ a non-negative integer) the HH operators are up to normalisation given by\footnote{\label{foot:delta}Note that these are indexed by a superscript in contrast to the \eqref{On}. These are the analytic continuation of $\delta_n(\tp)$ operators  \cite{Morris:2018mhd} under $a\cu\mapsto ia$.  See ref. \cite{Morris:2018mhd} for more properties.}
\be 
\label{HH}
\Op^n(\tp) := \text{e}^{a^2\tp^2} H_n(ia\tp)/(2ia)^n = \text{e}^{a^2\tp^2} \left( \tp^n +n(n-1)\,\tp^{n-2}/4a^2 +\cdots\right)\,, \qquad (\lambda=5+n) \,.
\ee
To prove this, note that after substituting $\tV(\tp) \cu\mapsto \tV(\tp)\, {\rm e}^{a^2\tp^2}$ into \eqref{eigen} and simplifying, one recovers the eigenoperator equation again but with $\lambda$ replaced by $9-\lambda$ and $a$ replaced with $ia$ \cite{Morris:2018mhd}. This new equation is therefore solved by \eqref{On} with $a$ replaced by $ia$, and $\lambda \cu=5\cu+n$.

The eigenoperator equation is of Sturm-Liouville  type. The corresponding Sturm-Liouville (SL) measure is 
\be 
\label{measure}
{\rm e}^{-a^2\tp^2}\,.
\ee
Perturbations $\tV(\tp)$ that are square integrable under this measure are particularly well behaved. We start by reviewing these properties. We then prove that there are solutions to the linearised flow equation \eqref{flow} that start inside this space, stay inside this space, and have all the desired properties that are usually taken for granted. In particular the linearised flow of the effective  interaction is then unique and can be split uniquely into a convergent sum over eigenoperators. The point of proving this is to contrast this with solutions that lie outside of this space. As we will see, the further we move outside of this space the less these properties can be taken for granted.

\section{SL perturbations}
\label{sec:squareint}

Let us refer to perturbations that are square integrable under the SL measure as ``SL perturbations'' and the space of such perturbations as the ``SL space''.\footnote{It has been studied within model approximations (which however are exact at the linearised level) in ref. \cite{Bridle:2016nsu}, see also \cite{Morris:1996nx,Morris:1996xq}, and exactly in ref. \cite{Morris:2018mhd}. In the following we also use insights from the behaviour for negative kinetic term, see already footnote \ref{foot:delta}. However we keep the exposition self-contained.}
Mathematically this space is a Hilbert space (which should not however be confused with state space in quantum mechanics). 
The eigenoperator solutions in the SL space are the Hermite polynomials \eqref{On}. In this space they are  orthonormal:
\be
\label{orthonormal}
\int^\infty_{-\infty}\!\!\!\! d\tp\,\, {\rm e}^{-a^2\tp^2} \cO{n}(\tp)\, \cO{m}(\tp) = \frac{1}{a}\left(\frac{1}{2a^2}\right)^n\! n!\sqrt{\pi}\,\delta_{nm}\,,
\ee
and complete. To see what this implies, start with some initial perturbation $V=V(\vp,\Lz)$ at an initial scale $\Lambda\cu=\Lz$. If $V(\vp,\Lz)$ grows slower than 
\be 
\label{boundi}
\frac1{\sqrt{|\vp|}}\exp\left( \frac{a^2\vp^2}{2\Lambda_0^2}\right)
\ee
as $\vp\cu\to\pm\infty$, then it is inside the SL space.  Then completeness means we are guaranteed a convergent expansion over the operators $\cO{n}$ in the square-integrable sense (which means it converges in the usual point-wise sense almost everywhere). Explicitly, if we define the coefficients
\be 
\label{tgni}
\tg_n(\Lz) = \frac{a}{\sqrt{\pi}}\frac{(2a^2)^n}{n!}\int^\infty_{-\infty}\!\!\!\! d\tp\,\, {\rm e}^{-a^2\tp^2} \cO{n}(\tp)\, \tV(\tp,\Lz)\,,
\ee
then we are guaranteed that the norm-squared of the remainder 
\be 
\label{completeness-proofi}
\int^\infty_{-\infty}\!\!\!\! d\tp\,\, {\rm e}^{-a^2\tp^2} \left( \tV(\tp,\Lz) - \sum_{n=0}^N \tg_n(\Lz)\, \cO{n}(\tp)\right)^{\!2}\to0\quad{\rm as}\quad N\to\infty\,,
\ee
vanishes as we send $N\cu\to\infty$. Taking the limit, we have a well-defined expansion 
\be 
\label{expansionscaledi}
\tV(\tp,\Lz) = \sum^\infty_{n=0} \tg_n(\Lz)\, \cO{n}(\tp)\,.
\ee
Notice that these coefficients $\tg_n(\Lz)$ are defined by \eqref{tgni} and are not yet expressed in terms of dimensionful couplings. However if we now define dimensionful couplings $g_n$ by writing
\be
\label{tgLambda}
\tg_n(\Lambda) \cu= g_n\, \Lambda^{n-4}\,,
\ee
(in particular at $\Lambda\cu=\Lz$) then the result builds in the separation of variables solution \eqref{lambda-v}. Thus the expansion
\be 
\label{expansionscaled}
\tV(\tp,\Lambda) = \sum^\infty_{n=0} \tg_n(\Lambda)\, \cO{n}(\tp)
\ee
provides the solution to the flow equation for the given initial condition $V=V(\vp,\Lz)$. Furthermore by orthonormality \eqref{orthonormal}, these $\tg_n(\Lambda)$ are also given by
\be 
\label{tgn}
\tg_n(\Lambda) = \frac{a}{\sqrt{\pi}}\frac{(2a^2)^n}{n!}\int^\infty_{-\infty}\!\!\!\! d\tp\,\, {\rm e}^{-a^2\tp^2} \cO{n}(\tp)\, \tV(\tp,\Lambda)\,,
\ee
at these lower scales. Finally \eqref{expansionscaled} still converges at these lower scales. To prove this we use the orthonormality relation \eqref{orthonormal} together with \eqref{tgLambda} to compute the norm-squared
\be 
\label{normsquared}
\int^\infty_{-\infty}\!\!\!\! d\tp\,\, {\rm e}^{-a^2\tp^2}\,  \tV^2(\tp,\Lambda) = \frac{\sqrt{\pi}}{\Lambda^8a} \sum_{n=0}^\infty n! \, g^2_n \left(\frac{\Lambda^2}{2a^2}\right)^n\,.
\ee
Since the RHS converges for $\Lambda\cu=\Lz$, the radius of convergence is greater than or equal to $\Lz$, and therefore it continues to converge for all $0\cu<\Lambda\cu\le\Lz$. Therefore we have proven that $\tV(\tp,\Lambda)$ remains in the SL space as $\Lambda$ is lowered. This in turn implies that the equivalent bound to \eqref{boundi} remains satisfied at lower scales, \ie for all $\Lambda\cu\le\Lz$ we have proven that
$V(\vp,\Lambda)$ grows slower than 
\be 
\label{bound}
\frac1{\sqrt{|\vp|}}\exp\left( \frac{a^2\vp^2}{2\Lambda^2}\right)
\ee
as $\vp\cu\to\pm\infty$.

Writing the expansion over eigenoperators \eqref{expansionscaled} instead in dimensionful terms
\be 
\label{expansionphys}
V(\vp,\Lambda) = \sum^\infty_{n=0} \Lambda^n g_n \cO{n}(\vp/\Lambda) = 
\sum^\infty_{n=0} g_n \Big(\vp^n -n(n-1)\Lambda^2\vp^{n-2}/4a^2 +\cdots\Big)\,,
\ee
the expansion converges for all $0\cu\le\Lambda\cu\le\Lz$, \ie also in the physical limit $\Lambda\cu\to0$. Notice that this means that the $g_n$ are just  the Taylor expansion coefficients of the physical potential $V(\vp,0)$ (at the linearised level):
\be 
\label{physical}
V(\vp,0) = \sum^\infty_{n=0} g_n\, \vp^n\,.
\ee
In fact $V(\vp,0)$ is an entire function, the RHS converging for all real $\vp$. This follows from \eqref{normsquared} since convergence of its RHS requires that the $g_n$ vanish faster than $1/\sqrt{n!}$ for large $n$. 

Substituting \eqref{tgn} into \eqref{expansionscaled} for the case $\tV(\tp,\Lz)=\tV_0(\tp)$, we have 
\be 
\label{greensol}
V(\vp,\Lambda) = \int^\infty_{-\infty}\!\!\!\! d\vp_0\,\, G_{\Lambda,\Lz}(\vp-\vp_0)\, V_0(\vp_0)\,,
\ee
where the Green's function $G$ is given by its spectral expansion:
\be 
\label{spectral}
G_{\Lambda,\Lz}(\vp-\vp_0) = \frac{a}{\Lz\sqrt{\pi}}\sum^\infty_{n=0}\frac1{n!}\left(\frac{2a^2\Lambda}{\Lz}\right)^n\!\cO{n}\!\left(\frac{\vp}{\Lambda}\right)\,\cO{n}\!\left(\frac{\vp_0}{\Lz}\right)\,.
\ee
We can get it in closed form by recognising that the flow equation \eqref{flow} is the heat diffusion equation in disguise. Indeed, introducing the `time' 
\be
\label{time}
T=\Lz^2-\Lambda^2\,, 
\ee
we get precisely the heat diffusion equation for diffusion coefficient $1/4a^2$:
\be 
\label{heat}
\frac{\partial}{\partial T}\, V(\vp,T) = \frac1{4a^2} \,V''(\vp,T)\,,
\ee
and thus from the well-known form of its Green's function (see \eg \cite{morse1953methods}) we find
\be 
\label{green}
G_{\Lambda,\Lz}(\vp-\vp_0) = \frac{a}{\sqrt{\pi(\Lambda^2_0-\Lambda^2)}}\,\exp \left(-\frac{a^2(\vp-\vp_0)^2}{\Lambda^2_0-\Lambda^2}\right)\,,\qquad (\Lambda<\Lz)\,.
\ee
Note that as a function of $\vp$ and $\Lambda$, $G_{\Lambda,\Lz}(\vp-\vp_0)$ satisfies the flow equation \eqref{flow} for $\Lambda\cu<\Lz$, as it must by \eqref{greensol} but which can also be checked explicitly. In fact it evidently satisfies the flow equation for all $\Lambda\cu\ne\Lz$. However,
since $G_{\Lambda,\Lz}(\vp-\vp_0)\cu\to\delta(\vp-\vp_0)$ as $\Lambda\cu\to\Lz$, and is pure imaginary for $\Lz\cu>\Lambda$, this representation only makes physical sense for $\Lambda\cu\le\Lz$, reflecting the fact that \eqref{heat} is parabolic, so that the Cauchy initial value problem is only well defined in the positive $T$ direction \ie for RG flows in the IR  direction. 

In fact while  the linearised RG flows can exist all the way to $\Lambda\to\infty$ (finite sums over the polynomials  \eqref{expansionscaled} such that $g_n=0$ for $n>n_\text{max}$ are examples), typically they 
extend only up to a finite range, failing at some higher critical scale. An example is provided by starting at $\Lambda\cu=\Lz$ with the bare potential 
\be 
\label{exV}
V_0(\vp)= 
A_1\,\exp \left(-\frac{\vp^2}{\mu_1^2}\right)+A_2\,\exp \left(-\frac{\vp^2}{\mu_2^2}\right)
\ee
where the $A_i$ are constants of dimension four and we set $0<\mu_1<\mu_2$. It is easy to see from \eqref{green} that this is $\sqrt{\pi}\mu_1 A_1\, G_{\Lz,\Lambda_1}(\vp)+\sqrt{\pi}\mu_2 A_2\, G_{\Lz,\Lambda_2}(\vp)$, where the $\Lambda_i \cu=\sqrt{\Lambda^2_0+a^2\mu^2_i}$. Therefore the solution to the flow equation is
\be
V(\vp,\Lambda) =  \sqrt{\pi}\mu_1 A_1\, G_{\Lambda,\Lambda_1}(\vp)+\sqrt{\pi}\mu_2 A_2\, G_{\Lambda,\Lambda_2}(\vp)\,,
\ee 
for all scales $\Lambda\cu<\Lambda_1$. However as $\Lambda$ approaches $\Lambda_1$ from below, 
\be V(\vp,\Lambda)\cu\to\sqrt{\pi}\mu_1 A\,  \delta(\vp)+\sqrt{\pi}\mu_2 A_2\, G_{\Lambda,\Lambda_2}(\vp)\,, \ee
after which the flow ceases to exist. If we persist in trying to use it for $\Lambda\cu>\Lambda_1$ we find a potential that is now complex.

This is also a generic feature. To see this, for simplicity assume a potential that is square integrable (such as is the case for \eqref{exV}). Then from the flow equation \eqref{flow} we see that
\be \Lambda\frac{\partial}{\partial\Lambda} \int^\infty_{-\infty}\!\!\!\!d\ph\, V^2(\ph,\Lambda) = \frac{\Lambda^2}{a^2}\int^\infty_{-\infty}\!\!\!\!d\ph \left\{ V'(\ph,\Lambda)\right\}^2\,, \ee
by integration by parts. Since the right hand side is positive, the integral over $V^2$ can only increase as $\Lambda$ increases, in turn increasing the right hand side even more. The integrals diverge at the singularity. The only way  the integral over $V^2$ can be finite once $\Lambda$ is above this, is if the right hand side then contributes an infinitely negative amount. But since the right hand side is the integral of a square this can only happen if $V$ is no longer be real.


%

\section{Alternative expansions}
\label{sec:alternatives}

We recall briefly the physical importance of being able to split the linearised flow uniquely into irrelevant, marginal, and relevant parts, as implied by \eqref{expansionscaled}. It leads to universality of the continuum limit since the latter is parametrised only by the few marginal and relevant operators \cite{Wilson:1973} (see \eg \cite{Morris:1998}).
However once we violate the bound \eqref{bound} (and are thus outside the SL space) it is no longer possible to split the linearised flow uniquely into relevant and irrelevant parts, at least in a way that holds for all points on the flow.
A perturbation can start outside the SL space, where it can be expanded in HH interactions \eqref{Kummers}, and then at lower scales it can enter the SL space where instead it is expanded in the polynomial eigenoperators. These two expansions can disagree about what parts of the flow are relevant.

For example consider the Green's function $G_{\Lambda,\Lz}(\vp)$ but where we replace $\Lz$ with $i\mu$. Since $G_{\Lambda,\Lz}(\vp)$ solves the flow equation when $\Lambda\cu\ne\Lz$, it also solves it for any $\mu\cu>0$. Discarding $\sqrt{\pi}$ and reintroducing the dimension four constant, $A$, we thus have the solution
\be 
\label{musol}
V(\vp,\Lambda) = \frac{\mu aA}{\sqrt{\mu^2+\Lambda^2}}\,\exp \left(\frac{a^2\vp^2}{\mu^2+\Lambda^2}\right)\,.
\ee
Notice that while $\Lambda>\mu$, this solution violates the bound \eqref{bound}.
Now we Taylor expand it in $\mu$, yielding odd powers of $\mu$, and thus by dimensions
\be 
\tV(\tp,\Lambda) = aA\sum^\infty_{n=0} \frac{\mu^{2n+1}}{\Lambda^{2n+5}}\, \tilde\upsilon_n(\tp)\,.
\ee
Since the RHS solves the flow equation for all $\mu$, it must be that the $\Lambda^{-2n-5}\tilde\upsilon_n(\tp)$ separately solve the flow equation. We recognise (from both the eigenvalue and the asymptotic behaviour) that these are proportional to the HH interactions defined in \eqref{HH}, and therefore
\be 
\label{HHexpansion}
V(\vp,\Lambda) = A \sum^\infty_{n=0} c_n \left({a\mu/\Lambda}\right)^{2n+1}\Op^{2n}\!\left({\vp}/{\Lambda}\right)\,,
\ee
where the $c_n$ are numbers. In complex $\mu$ space, \eqref{musol} is analytic except at $\mu\cu=\pm i\Lambda$. Thus the series above converges for all $\vp$ and for all $\Lambda\cu>\mu$.

On the other hand when $\Lambda\cu<\mu$ , we see that the growth for large $\vp$ lies inside the bound \eqref{bound}, and thus that the solution lies inside the SL space. Therefore here we can write the solution \eqref{musol} uniquely as a series expansion over the polynomial  eigenoperators $\cO{n}(\tp)$. Recalling \eqref{physical}, we can read off the conjugate couplings from the Taylor expansion of $V(\vp,0)$. Thus using \eqref{expansionphys}, we have the expansion
\be 
\label{OnExpansion}
V(\vp,\Lambda) = aA\sum^\infty_{n=0} \frac1{n!}\left(\frac{a\Lambda}{\mu}\right)^{2n}\!\!\cO{2n}\!\left(\frac{\vp}{\Lambda}\right)\,,
\ee
which converges for all $\vp$ and for all $\Lambda<\mu$. (The expansion can also be derived directly from \eqref{spectral} on using properties of Hermite polynomials.)

We thus have two equivalent descriptions of the same flow but with apparently contradictory RG behaviour. In the former case \eqref{HHexpansion}, one would deduce that the flow involves only relevant couplings, since in the expansion all the eigenoperators are relevant. In the latter case \eqref{OnExpansion} however, all the eigenoperators are irrelevant apart from the first three.

\section{Singular flows}
\label{sec:singular}

We have already seen at the end of sec. \ref{sec:squareint} that flows upwards can end in a singularity. This is not unexpected: there is no reason \textit{a priori} why a freely chosen effective action $V(\vp,\Lz)$ should be the result of integrating out modes starting with some potential $V(\vp,\Lambda_1)$ at a higher scale $\Lambda_1\cu>\Lz$. However once we are even further outside the SL space, it is also the case that linearised flows towards the IR may end in a singularity. This can happen if $V(\ph,\Lambda)$ grows faster than the \emph{square} of the bound \eqref{bound} (as we explain in the next section).
The solution
\be 
\label{singular}
V(\vp,\Lambda) = \frac{\mu aA}{\sqrt{\Lambda^2-\mu^2}}\,\exp \left(\frac{a^2\vp^2}{\Lambda^2-\mu^2}\right)\,, \qquad (\Lambda>\mu)\,,
\ee
will illustrate this behaviour. It is just the previous solution \eqref{musol} after replacing $\mu$ with $i\mu$, and dividing by $i$. 
Comparing to the bound \eqref{bound}, we see that \eqref{singular} indeed grows faster than its square (for all $\Lambda\cu>\mu$). Mapping from \eqref{HHexpansion}, we see that it has a convergent HH expansion in this domain:
\be 
\label{HHexpSingular}
V(\vp,\Lambda) =  \sum^\infty_{n=0} (-)^nc_n \left({a\mu/\Lambda}\right)^{2n+1}\Op^{2n}\!\left({\vp}/{\Lambda}\right)\,,
\ee
(the $c_n$ being the same numbers as before).
However we see from \eqref{singular} that it ends in a singularity as $\Lambda\cu\to\mu$, where it diverges for all values of $\vp$. To be clear we emphasise that since it diverges everywhere, there is no sense in which it can still be regarded as a solution once we reach $\Lambda\cu=\mu$. If we nevertheless cavalierly attempt to continue the solution below this point, the solution becomes pure imaginary. 

Notice that \eqref{singular} is proportional to the Green's function $G_{\Lambda,\mu}(\vp)$ continued above its domain of validity. A generic solution illustrating these properties would follow from the Green's function representation \eqref{greensol} if we choose $V_0(\vp_0)$ to be integrable, imaginary and of compact support. Then
\be 
V(\vp,\Lambda) = \int^\infty_{-\infty}\!\!\!\! d\vp_0\,\, G_{\Lambda,\mu}(\vp-\vp_0)\, V_0(\vp_0)\,,
\ee
is a solution that is real and well behaved for $\Lambda\cu>\mu$ but which diverges everywhere as $\Lambda\cu\to\mu$ from above.

\section{Non-unique flows}
\label{sec:nonunique}

In the Wilsonian RG literature it is taken for granted that the solution to the flow equation is unique once the initial effective action is specified: $V(\vp,\Lz) \cu= V_0(\vp)$. 
In fact this property is true for solutions only if they grow sufficiently slowly for large $\vp$. 

Of course by construction the solution expanded over the eigenoperators $\cO{n}(\tp)$  (\ref{expansionscaled},\ref{expansionphys}), or  written as the convolution \eqref{greensol}, is unique. While the former makes sense only if the initial perturbation $V_0(\vp)$ grows slower than the bound \eqref{boundi}, the latter  converges for a larger space of initial perturbations $V_0(\vp)$. From the explicit form of the Green's function \eqref{green}, the convolution form of the solution converges for all $\Lambda\le\cu\Lz\,$, provided $V_0(\vp)$ is integrable and grows slower than the square of the bound \eqref{boundi}. 

Note that these properties are consistent with our previous examples. Example \eqref{musol} lies outside the bound \eqref{bound} for $\Lambda\cu>\mu$ and therefore does not have an expansion over the $\cO{n}$ there. However it lies inside the square the bound. Therefore it has a convergent Green's function representation which gives a non-singular flow to the IR. On the other hand for any $\Lambda\cu>\mu$, the example \eqref{singular} violates both the bound \eqref{bound} and its square. Therefore it can neither be expanded over the $\cO{n}$ nor does it have a convergent Green's function representation. This is reflected in the fact that actually the solution becomes singular as $\Lambda\cu\to\mu$ from above.

Even though the Green's function construction, when convergent, yields a unique solution,
this does not mean it is the only solution. It is however the only solution if we restrict the solution space to $V(\vp,\Lambda)$ that grow slower than \emph{some} exponential of $\vp^2$ \ie to solutions that, for all $0\le \Lambda\le\Lambda_0$, grow slower than
\be 
\label{Bound}
\exp( B\vp^2)\,, 
\ee
for some fixed (sufficiently large) positive constant $B$. The proof follows from the equivalence (\ref{time},\ref{heat}) to the heat equation,  since uniqueness of such bounded solutions is proven for the latter. See \eg Theorem 7 of sec. 2.3 in ref. \cite{evans10} where the proof is the result of applying the maximum principle together with some careful limits. 

\begin{figure}[ht]
\centering
 \includegraphics[width=0.45\textwidth]{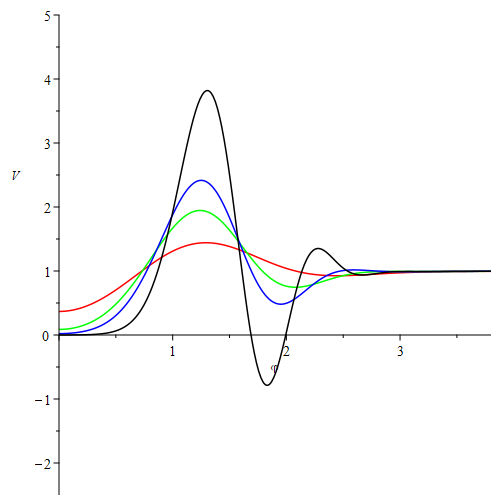}
 \includegraphics[width=0.45\textwidth]{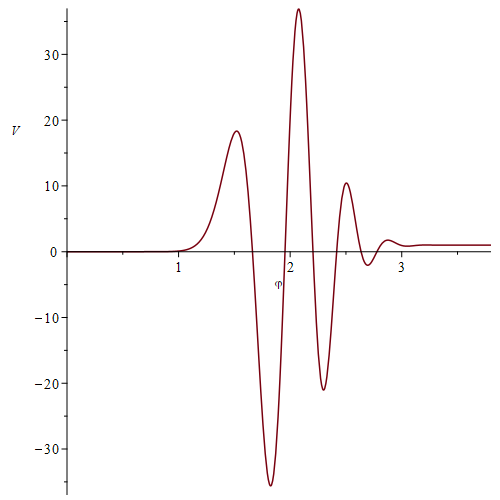} 
 \caption{The $\alpha=2$ Tychonoff solution in units of $v$ and $\mu/a$. On the left, it is plotted for $\Lambda=0$ (red), $0.6\,\mu$ (green), $0.7\,\mu$ (blue) and $0.8\,\mu$ (black), and on the right it is plotted for $\Lambda=0.9\,\mu$.}
 \label{fig:wiggle}
\end{figure}

Now, we show that solutions are no longer unique if they are allowed to grow faster than any such exponential \eqref{Bound}. In this case one can have two solutions $V_1(\vp,\Lambda)$ and $V_2(\vp,\Lambda)$ of the flow equation \eqref{flow} which agree for all scales $\Lambda\ge\mu$, but disagree once $\Lambda\cu<\mu$. 
Since the flow equation is linear, this is equivalent to the statement that their difference, $V(\vp,\Lambda)=V_1(\vp,\Lambda)-V_2(\vp,\Lambda)$, is a non-trivial solution that nevertheless vanishes identically at all scales $\Lambda\cu\ge\mu$. Using the equivalence (\ref{time},\ref{heat}) and following Tychonoff \cite{Tyc35}, we show that
\be \label{tyc}
V(\ph,\Lambda) = v\sum_{k=0}^\infty \frac1{(2k)!}\left(\frac{2a\ph}{\mu}\right)^{2k}\!g^{(k)}\!\left(\frac{\mu^2-\Lambda^2}{\mu^2}\right)\ee
is such a solution ($v$ being a proportionality constant of dimension four).
It is constructed from the $k^\text{th}$ derivatives of the function 
\beal g(t) &= \mathrm{e}^{-t^{-\alpha}}\qquad\,\text{for}\quad  t>0\nn\\
&= 0 \qquad\qquad\text{for}\quad   t\le0
\eeal
where one must choose the parameter $\alpha>1$. Note that $g(t)$ and all its derivatives are continuous at $t=0$, so in the series \eqref{tyc} each term  vanishes smoothly as $\Lambda\cu\to\mu$ from below and of course vanishes identically for all $\Lambda\cu\ge\mu$. It is straightforward to verify by direct substitution that \eqref{tyc} is indeed a (formal) solution of the flow equation \eqref{flow}. It would only be a formal solution however unless one can show that the series \eqref{tyc} converges. In fact the series is absolutely convergent, \cf app. \ref{app:envelope} \cite{Tyc35,john1991partial}. From that analysis one also sees that $V$ stays within the envelopes
\be \label{tycb} \left| V(\ph,\Lambda)\right| \le |v|
\exp\left[\frac{4a^2\ph^2}{(\mu^2-\Lambda^2)r}-F(r)\left(\frac{\mu^2}{\mu^2-\Lambda^2}\right)^{\!\alpha}\,\right]\qquad\text{for}\quad  \Lambda<\mu\,.\ee
These envelopes are parametrised by an $0\cu<r\cu<1$ which additionally must be chosen so that 
\be 
\label{Refactor}
\mathrm{Re}\left(1+r\,\mathrm{e}^{i\theta}\right)^{-\alpha}
\ee
is bounded below by a positive number $F(r)$
for all $0\le\theta<2\pi$ (see app. \ref{app:envelope}). Note that these envelopes all exceed the bound \eqref{Bound} for $\Lambda$ sufficiently close to $\mu$ (whatever we choose for $B$) but, since $\alpha\cu>1$, for any fixed $\ph$ they vanish  as $\mu\to\Lambda$. This demonstrates that $V(\ph,\Lambda)$ also vanishes as $\mu\to\Lambda$ (in fact uniformly for bounded complex $\ph$).

As $\Lambda$ is lowered through the critical point $\Lambda\cu=\mu$, the Tychonoff solution \eqref{tyc} takes the form of a divergent wave-packet that comes in from infinite $\ph$, see fig. \ref{fig:wiggle}. Lowering $\Lambda$ still further, the solution becomes less oscillatory. At values of $\ph$ much less than the wave-packet position we still have $V\approx0$, while for values $\ph$ much larger than the wave-packet position $V$ rapidly tends to $v$.

(The fact that the series in \eqref{tyc} tends to $1$ as $\ph\to\infty$ seems far from obvious but can be convincingly demonstrated numerically. Individual terms in \eqref{tyc} become very large but cancel each other to a high degree. For example for the point $\ph=3.8\,\mu/a$ in the right hand plot fig. \ref{fig:wiggle}, individual terms grow to $10^{45}\,v$. To get accurate results required high digits accuracy and many terms, \eg 90 digits and close to 300 terms for the right hand plot.
Numerically we established that $V\to v$ for large $\ph$ by working to even greater accuracy. For example for $\Lambda\cu=0.6\,\mu$ we followed the solution out to $\ph=16$ and established that there $V=0.9999\,v$ to four decimal places. This required however working to 244 digits accuracy and summing $1230$ terms.)

\section{Conclusions}
\label{sec:conclusions}

We summarise our main findings. Non-singular solutions $V(\ph,\Lambda)$ to the linearised flow equation \eqref{flow}, that grow slower than \eqref{bound}:
\be \frac1{\sqrt{|\vp|}}\exp\left( \frac{a^2\vp^2}{2\Lambda^2}\right)\,, \label{boundrepeat}\ee
are square-integrable under the Sturm-Liouville measure. They can be expanded over polynomial eigenoperators, the Hermite polynomials \eqref{On}, with the series converging in the square integrable sense. The flow towards the IR is unique and non-singular. Generically flows towards the ultraviolet however fail at a singularity, after which the solution no longer exists, at least as a real solution. We proved this in sec. \ref{sec:squareint}. 

The uniqueness of the expansion over eigenoperators is an important property since it allows for the universality of the continuum limit, this being parametrised by the marginal/relevant couplings that can be uniquely identified in this expansion. However if solutions grow at large $\ph$ in a way such as to exceed the above bound, they need no longer have a unique expansion over eigenoperators. In sec. \ref{sec:alternatives} we demonstrated this by the solution \eqref{musol}. It has a convergent expansion over the HH eigenoperators \eqref{HH} for $\Lambda>\mu$ such that all are relevant, but has a convergent expansion over the polynomial eigenoperators \eqref{On} when $\Lambda<\mu$ such that all but three of them are irrelevant.

If the solution grows faster than the square of the above bound, the flow towards the IR can end in a singularity and thus lead to flows that cannot be completed (\ie such that there is an obstruction to integrating out all the modes). We saw this in sec. \ref{sec:singular}, where we also saw that a convergent expansion over HH eigenoperators can lead to such singularities and thus incomplete flows. In sec. \ref{sec:nonunique} we related this to the fact that solutions that grow faster than the square of the above, no longer have a convergent Green's function representation.

Finally in sec. \ref{sec:nonunique} we saw that if we allow growth faster than any exponential $\mathrm{e}^{B\ph^2}$ (with fixed constant $B$) then solutions are no longer uniquely determined by the initial `bare' potential $V(\ph,\Lz)$. New interactions can spontaneously appear at lower scales through `Tychonoff' wave-packets that travel in from $\ph=\infty$. 

It is tempting to try and find a physical r\^ole for such an effect, just as it was for the HH eigenoperators \cite{HHOrig,HHReply,HH2ndpaper,Percacci:2003jz,Periwal:1995hw,Halpern:1997gn,Branchina:2000jp,Bonanno:2000sy,Gies:2000xr,Altschul:2004yq,Altschul:2004gt,Altschul:2005mu,Gies:2009hq,Huang:2010qn,Huang:2011xg,Huang:2011xha,Pietrykowski:2012nc,Huang:2013zaa,Abhignan:2021lub,Morley:2018omd}  (see our discussion in sec. \ref{sec:intro}), and indeed it is tempting to search for physical meaning in the other challenging effects we have just summarised. However our objections \cite{Morris:1996nx,Morris:1996xq,Bridle:2016nsu} to HH interactions apply equally well to all these effects. In particular  if we use the flow equation for the Legendre effective action with IR cutoff
\cite{Nicoll1977,Wetterich:1992,Morris:1993}
\be
\label{Legendre}
{\partial\over\partial\Lambda} {\Gamma}[\ph] = {1\over2} {\rm tr}  \left[ {\cal R} + {\delta^2{\Gamma}\over\delta\ph\delta\ph}\right]^{-1} {\partial {\cal R}\over\partial\Lambda}\,,
\ee
then for any interaction that exceeds \eqref{boundrepeat}, the right hand side is forced to vanish at large $\ph$, no matter how small we make the interaction at any finite $\ph$ \cite{Bridle:2016nsu}. It follows that working with the linearised flow equation \eqref{flow} is not  justified at large  $\ph$. Instead at sufficiently large $\ph$, the flow equation collapses to $\partial_\Lambda{\Gamma}[\ph]=0$, \ie mean-field evolution takes over such that in fact the action is frozen out and independent of $\Lambda$. For interactions that grow at large field like $\mathrm{e}^{B\ph^2}$ for some $B$, since they are frozen out, we find that at scales $\Lambda<a/\sqrt{2B}$ they will again be inside the bound \eqref{boundrepeat} and thus have `fallen' back into the SL space where they can be expanded over the polynomial eigenoperators as \eqref{OnExpansion} \cite{Morris:1996nx,Morris:1996xq,Bridle:2016nsu}. Interactions that grow faster than $\mathrm{e}^{B\ph^2}$ for any $B$ (for example $V\sim\mathrm{e}^{C\ph^4}$) do not fall back into the SL space, but nevertheless  their exact evolution is very different from that described by the linearised flow equation \eqref{flow}.

\section*{Acknowledgments}

TRM acknowledges support from STFC through Consolidated Grant ST/T000775/1.


\appendix
\section{Convergence and envelopes}
\label{app:envelope}

Here we derive the envelope formula \eqref{tycb} and prove absolute convergence of the series \eqref{tyc}. Our derivation closely follows the exposition of Tychonoff's proof as given in chapt. 7 of \cite{john1991partial}. 

First we note that the series \eqref{tyc} is absolutely convergent if the sum 
\be \label{sumabs}  \sum_{k=0}^\infty \frac1{(2k)!}\left|\left(\frac{2a\ph}{\mu}\right)^{2k}\!g^{(k)}\!\left(\frac{\mu^2-\Lambda^2}{\mu^2}\right) \right|\,, \ee
converges, \ie where all terms are taken positive. To prove that the above sum converges, we use Cauchy's representation for derivatives of analytic functions
\be\label{CauchyCont} g^{(k)}(t) = \frac{k!}{2\pi i}\oint\! dz\, \frac{\mathrm{e}^{-z^{-\alpha}}}{(z-t)^{k+1}}\,,  \ee
where we take real $t\cu>0$, putting us in the regime $\Lambda\cu<\mu$. Choosing the contour to be the circle $z= t(1+r\,\mathrm{e}^{i\theta})$, where $0\le\theta<2\pi$ and $r$ is fixed in the range $0\cu<r\cu<1$, we have 
\be 
\mathrm{Re}(-z^{-\alpha})= -t^{-\alpha}\,\mathrm{Re} \left(1+r\,\mathrm{e}^{i\theta}\right)^{-\alpha}\,.
\ee
Now if $r$ is small enough the last factor, which is \eqref{Refactor}, is bounded below by a positive constant, which we call $F(r)$. This can be determined by minimising over $\theta$.
For example if $\alpha=2$ we find that we must have $r<\frac1{\sqrt{2}}$. Then we find that $F(r)=\frac12(1-2r^2)/(1-r^2)^2$ for $r>\frac12$, while for $0<r<\frac12$ we have $F(r)=1/(1+r)^2$. Thus from \eqref{CauchyCont} we have that
\be 
\left| g^{(k)}(t)\right| \le \frac{k!}{(rt)^k}\,\mathrm{e}^{-F(r)t^{-\alpha}}\,.
\ee
Finally, since $k!/(2k)!<1/k!$, we see that the $k^\text{th}$ term in \eqref{sumabs} is bounded above by
\be \frac1{k!} \left(\frac{4a^2\ph^2}{(\mu^2-\Lambda^2)r}\right)^{k}\,\mathrm{e}^{-F(r)t^{-\alpha}}\,.\ee
Since the sum of these terms converges, it follows that the sum \eqref{sumabs} converges, and thus that the sum in \eqref{tyc} is absolutely convergent. In fact the above is just a term-wise expansion of the exponential in \eqref{tycb} (up to the factor $|v|$). Since its sum is larger than \eqref{sumabs} which in turn is larger in magnitude than the sum in \eqref{tyc}, we have also proven that $|V(\ph,\Lambda)|$ is bounded by the envelopes \eqref{tycb}.

\vfill
\newpage 

\bibliographystyle{hunsrt}
\bibliography{references} 

\begin{thebibliography}{10}

\bibitem{HHOrig}
Kenneth Halpern and Kerson Huang.
\newblock Fixed-point structure of scalar fields.
\newblock {\em Phys. Rev. Lett.}, 74:3526--3529, May 1995.

\bibitem{HHReply}
Kenneth Halpern and Kerson Huang.
\newblock Halpern and huang reply:.
\newblock {\em Phys. Rev. Lett.}, 77:1659--1659, Aug 1996.

\bibitem{HH2ndpaper}
Kenneth Halpern and Kerson Huang.
\newblock Nontrivial directions for scalar fields.
\newblock {\em Phys. Rev. D}, 53:3252--3259, Mar 1996.

\bibitem{Percacci:2003jz}
Roberto Percacci and Daniele Perini.
\newblock Asymptotic safety of gravity coupled to matter.
\newblock {\em Phys.Rev.}, D68:044018, 2003, hep-th/0304222.

\bibitem{Periwal:1995hw}
Vipul Periwal.
\newblock Halpern-huang directions in effective scalar field theory.
\newblock {\em Mod.Phys.Lett.}, A11:2915--2920, 1996, hep-th/9512108.

\bibitem{Halpern:1997gn}
Kenneth Halpern.
\newblock Cross-section and effective potential in asymptotically free scalar
  field theories.
\newblock {\em Phys.Rev.}, D57:6337--6341, 1998, hep-th/9708124.

\bibitem{Branchina:2000jp}
Vincenzo Branchina.
\newblock Nonperturbative renormalization group potentials and quintessence.
\newblock {\em Phys.Rev.}, D64:043513, 2001, hep-ph/0002013.

\bibitem{Bonanno:2000sy}
Alfio Bonanno.
\newblock Nonperturbative scaling in the scalar theory.
\newblock {\em Phys.Rev.}, D62:027701, 2000, hep-th/0001060.

\bibitem{Gies:2000xr}
Holger Gies.
\newblock Flow equation for halpern-huang directions of scalar o(n) models.
\newblock {\em Phys.Rev.}, D63:065011, 2001, hep-th/0009041.

\bibitem{Altschul:2004yq}
B.~Altschul.
\newblock Nonpolynomial normal modes of the renormalization group in the
  presence of a constant vector potential background.
\newblock {\em Nucl.Phys.}, B705:593--604, 2005, hep-th/0403093.

\bibitem{Altschul:2004gt}
B.~Altschul.
\newblock Asymptotically free lorentz and cpt-violating scalar field theories.
\newblock 2004, hep-th/0407173.

\bibitem{Altschul:2005mu}
B.~Altschul and V.~Alan Kostelecky.
\newblock Spontaneous lorentz violation and nonpolynomial interactions.
\newblock {\em Phys.Lett.}, B628:106--112, 2005, hep-th/0509068.

\bibitem{Gies:2009hq}
Holger Gies and Michael~M. Scherer.
\newblock Asymptotic safety of simple yukawa systems.
\newblock {\em Eur.Phys.J.}, C66:387--402, 2010, 0901.2459.

\bibitem{Huang:2010qn}
Kerson Huang, Hwee-Boon Low, and Roh-Suan Tung.
\newblock Cosmology of an asymptotically free scalar field with spontaneous
  symmetry breaking.
\newblock 2010, 1011.4012.

\bibitem{Huang:2011xg}
Kerson Huang, Hwee-Boon Low, and Roh-Suan Tung.
\newblock Scalar field cosmology i: Asymptotic freedom and the initial-value
  problem.
\newblock {\em Class.Quant.Grav.}, 29:155014, 2012, 1106.5282.

\bibitem{Huang:2011xha}
Kerson Huang, Hwee-Boon Low, and Roh-Suan Tung.
\newblock Scalar field cosmology ii: Superfluidity and quantum turbulence.
\newblock {\em Int.J.Mod.Phys.}, A27:1250154, 2012, 1106.5283.

\bibitem{Pietrykowski:2012nc}
Artur~R. Pietrykowski.
\newblock Interacting scalar fields in the context of effective quantum
  gravity.
\newblock {\em Phys.Rev.}, D87:024026, 2013, 1210.0507.

\bibitem{Huang:2013zaa}
Kerson Huang.
\newblock A critical history of renormalization.
\newblock {\em Int. J. Mod. Phys. A,}, 28:1330050, 2013, 1310.5533.

\bibitem{Abhignan:2021lub}
Venkat Abhignan and R.~Sankaranarayanan.
\newblock {Casimir-Like Effect from Thermal Field Fluctuations}.
\newblock {\em Braz. J. Phys.}, 51(6):1897--1903, 2021.

\bibitem{Morley:2018omd}
Peter~D. Morley.
\newblock {Renormalizable Gravitational Action That Reduces to General
  Relativity on the Mass-Shell}.
\newblock {\em Galaxies}, 6(3):81, 2018.

\bibitem{Morris:1996nx}
Tim~R. Morris.
\newblock {On the fixed point structure of scalar fields}.
\newblock {\em Phys. Rev. Lett.}, 77:1658, 1996, hep-th/9601128.

\bibitem{Morris:1996xq}
Tim~R. Morris.
\newblock {Three-dimensional massive scalar field theory and the derivative
  expansion of the renormalization group}.
\newblock {\em Nucl.Phys.}, B495:477--504, 1997, hep-th/9612117.

\bibitem{Bridle:2016nsu}
I.~Hamzaan~Bridle and Tim~R. Morris.
\newblock {Fate of nonpolynomial interactions in scalar field theory}.
\newblock {\em Phys. Rev.}, D94:065040, 2016, 1605.06075.

\bibitem{Morris:1994jc}
Tim~R. Morris.
\newblock {The Renormalization group and two-dimensional multicritical
  effective scalar field theory}.
\newblock {\em Phys.Lett.}, B345:139--148, 1995, hep-th/9410141.

\bibitem{Wilson:1973}
K.G. Wilson and John~B. Kogut.
\newblock {The Renormalization group and the epsilon expansion}.
\newblock {\em Phys.Rept.}, 12:75--200, 1974.

\bibitem{Wegner:1972ih}
Franz~J. Wegner and Anthony Houghton.
\newblock {Renormalization group equation for critical phenomena}.
\newblock {\em Phys. Rev.}, A8:401--412, 1973.

\bibitem{Polchinski:1983gv}
Joseph Polchinski.
\newblock {Renormalization and Effective Lagrangians}.
\newblock {\em Nucl.Phys.}, B231:269--295, 1984.

\bibitem{Nicoll1977}
J.~F. Nicoll and T.~S. Chang.
\newblock {An Exact One Particle Irreducible Renormalization Group Generator
  for Critical Phenomena}.
\newblock {\em Phys. Lett.}, A62:287--289, 1977.

\bibitem{Wetterich:1992}
Christof Wetterich.
\newblock {Exact evolution equation for the effective potential}.
\newblock {\em Phys.Lett.}, B301:90--94, 1993.

\bibitem{Morris:1993}
Tim~R. Morris.
\newblock {The Exact renormalization group and approximate solutions}.
\newblock {\em Int.J.Mod.Phys.}, A 09:2411--2450, 1994, hep-ph/9308265.

\bibitem{Morris:2018mhd}
Tim~R. Morris.
\newblock {Renormalization group properties in the conformal sector: towards
  perturbatively renormalizable quantum gravity}.
\newblock {\em JHEP}, 08:024, 2018, 1802.04281.

\bibitem{Manrique:2008zw}
Elisa Manrique and Martin Reuter.
\newblock {Bare Action and Regularized Functional Integral of Asymptotically
  Safe Quantum Gravity}.
\newblock {\em Phys.Rev.}, D79:025008, {(2009)}, 0811.3888.

\bibitem{Morris:2015oca}
Tim~R. Morris and Zo{\"e}~H. Slade.
\newblock {Solutions to the reconstruction problem in asymptotic safety}.
\newblock {\em JHEP}, 11:094, 2015, 1507.08657.

\bibitem{Nicoll:1974zz}
J.~F. Nicoll, T.~S. Chang, and H.~E. Stanley.
\newblock {Approximate Renormalization Group Based on the Wegner-Houghton
  Differential Generator}.
\newblock {\em Phys. Rev. Lett.}, 33:540--543, 1974.

\bibitem{Morris:1998}
Tim~R. Morris.
\newblock {Elements of the continuous renormalization group}.
\newblock {\em Prog.Theor.Phys.Suppl.}, 131:395--414, 1998, hep-th/9802039.

\bibitem{Aoki:2000wm}
K.~Aoki.
\newblock {Introduction to the nonperturbative renormalization group and its
  recent applications}.
\newblock {\em Int. J. Mod. Phys.}, B14:1249--1326, 2000.

\bibitem{Bagnuls:2000}
C.~Bagnuls and C.~Bervillier.
\newblock {Exact renormalization group equations. An Introductory review}.
\newblock {\em Phys.Rept.}, 348:91, 2001, hep-th/0002034.

\bibitem{Berges:2000ew}
Juergen Berges, Nikolaos Tetradis, and Christof Wetterich.
\newblock {Nonperturbative renormalization flow in quantum field theory and
  statistical physics}.
\newblock {\em Phys. Rept.}, 363:223--386, 2002, hep-ph/0005122.

\bibitem{Polonyi:2001se}
Janos Polonyi.
\newblock {Lectures on the functional renormalization group method}.
\newblock {\em Central Eur. J. Phys.}, 1:1--71, 2003, hep-th/0110026.

\bibitem{Pawlowski:2005xe}
Jan~M. Pawlowski.
\newblock {Aspects of the functional renormalisation group}.
\newblock {\em Annals Phys.}, 322:2831--2915, 2007, hep-th/0512261.

\bibitem{Delamotte:2007pf}
Bertrand Delamotte.
\newblock {An Introduction to the nonperturbative renormalization group}.
\newblock {\em Lect. Notes Phys.}, 852:49--132, 2012, cond-mat/0702365.

\bibitem{Kopietz:2010zz}
Peter Kopietz, Lorenz Bartosch, and Florian Schutz.
\newblock {Introduction to the functional renormalization group}.
\newblock {\em Lect. Notes Phys.}, 798:1--380, 2010.

\bibitem{Rosten:2010vm}
Oliver~J. Rosten.
\newblock {Fundamentals of the Exact Renormalization Group}.
\newblock {\em Phys. Rept.}, 511:177--272, 2012, 1003.1366.

\bibitem{Dupuis:2020fhh}
N.~Dupuis, L.~Canet, A.~Eichhorn, W.~Metzner, J.~M. Pawlowski, M.~Tissier, and
  N.~Wschebor.
\newblock {The nonperturbative functional renormalization group and its
  applications}.
\newblock 2020, 2006.04853.

\bibitem{AbramowitzStegun}
Milton Abramowitz and Irene~A. Stegun.
\newblock {\em Handbook of Mathematical Functions with Formulas, Graphs, and
  Mathematical Tables}.
\newblock Dover, New York, 9th edition, 1964.

\bibitem{OlverAsymptotics}
F.~W.~J. Olver.
\newblock {\em Asymptotics and special functions}.
\newblock Academic Press, New York-London, 1974.
\newblock Computer Science and Applied Mathematics.

\bibitem{LebedevSpecialFunctions}
N.~N. Lebedev.
\newblock {\em Special Functions and Their Applications}.
\newblock Dover Publications, New York, 1972.
\newblock Originally published by Prentice-Hall in 1965.

\bibitem{morse1953methods}
P.M.C. Morse and H.~Feshbach.
\newblock {\em Methods of Theoretical Physics}.
\newblock International series in pure and applied physics. McGraw-Hill, 1953.

\bibitem{evans10}
Lawrence~C. Evans.
\newblock {\em Partial differential equations}.
\newblock American Mathematical Society, Providence, R.I., 2010.

\bibitem{Tyc35}
A~Tychonoff.
\newblock Th{\'e}or{\`e}mes d'unicit{\'e} pour l'{\'e}quation de la chaleur.
\newblock {\em Mat. Sb.}, 42(2):199, 1935.

\bibitem{john1991partial}
F.~John.
\newblock {\em Partial Differential Equations}.
\newblock Applied Mathematical Sciences. Springer New York, 1991.

\end{thebibliography}

\end{document}